\documentstyle[11pt,newpasp,twoside,epsf]{article}
\newcommand{\w}{$W_0^{\lambda2796}$}

\begin{document}
\title{The Evolution of Neutral Gas in the Universe as Traced by
Damped Lyman Alpha Systems}
\author{Sandhya M. Rao and David A. Turnshek}
\affil{Department of Physics and Astronomy, University of Pittsburgh, 
Pittsburgh, PA 15260}

\begin{abstract}

We discuss our recent results on the  statistical properties of damped
Lyman alpha systems (DLAs) at low redshift ($z<1.65$) (Rao \& Turnshek
2000). Contrary  to expectations, we found that the cosmological
neutral gas  mass density as traced by DLAs, $\Omega_{DLA}$, does not
evolve from redshifts $z\approx 4$ to $z\approx 0.5$ and that
extrapolation to $z=0$ results in a value that is  a factor of $\sim6.5$
times higher than what is derived from  galaxies at the current epoch
using HI 21 cm emission  measurements.  We review the current status
of HI measurements at low redshift and at the current epoch,
and discuss possible causes of this discrepancy.

\end{abstract} 

\section{Introduction}

It was recognized in the mid-eighties that damped Lyman alpha systems
(DLAs) reveal unique information about the evolutionary history  of
the Universe. Optical surveys of QSOs showed that DLAs trace the bulk
of the observable neutral gas mass at high redshift, and therefore,
that they can be used to study the formation and evolution of
galaxies.  A particularly compelling result was that the comoving
neutral gas  mass density in DLAs at redshift $z\approx 3.5$ was
comparable  to the luminous mass density observed in galaxies at the
current epoch.  There also seemed to be evidence for a decline in the
neutral gas mass density from $z\approx3.5$ to $z\approx1.7$ that
extrapolated to the value inferred  from gas-rich galaxies at $z=0$
(Wolfe et al. 1986; Turnshek et al.  1989; Lanzetta et al. 1991; Rao
\& Briggs 1993). This was interpreted  as evidence for gas depletion
due to star formation, and DLAs were thought to trace the evolution of
the progenitors of present-day disk galaxies.

However, several studies carried out during the second half of the
last decade have shown that the interpretation is not so
straightforward.  Groundbased surveys can search for DLA lines only
beyond $z=1.65$ when the Ly$\alpha$ line moves into the optical region
of the electromagnetic spectrum. Thus, until the recent IUE and HST
surveys for DLAs (Lanzetta, Wolfe, \& Turnshek 1995; Rao, Turnshek, \&
Briggs 1995; Jannuzi et al. 1998; Rao \& Turnshek 2000, henceforth,
RT2000), about 70\% of the Ly$\alpha$ Universe was unexplored.  The
RT2000 survey, which we discuss in this contribution, more than
doubled the number of known low-redshift DLAs. We used a bootstrapping
method which relied on the statistics of MgII absorption lines to
identify DLAs.  The results, although still limited by small number
statistics, show that $\Omega_{DLA}$ is consistent with remaining
constant from redshifts $z\approx 4$ to $z\approx 0.5$ and that
extrapolation to $z=0$ results in a value that is a factor of
$\sim6.5$ times higher than what is derived from galaxies at the
current epoch.  The DLA redshift number density is only
marginally consistent with  evolution between redshifts $z\approx4$ and
$z\approx0.5$. This implies that DLAs may be a slowly evolving
population that do not undergo significant gas depletion and, by
extension, star formation.

Apart from the low-redshift DLA statistics, results from metallicity
studies, kinematics, and direct imaging of the DLAs have shown that
the ``DLA - disk galaxy'' paradigm is inadequate.  Pettini and
collaborators (e.g. Pettini et al.  1997, Pettini et al. 1999, Pettini
et al.  2000) have shown that there is no obvious tendency for DLA
metallicities to increase with decreasing redshift and approach the
solar value.  However, there is considerable scatter in the individual
metallicities which is most likely caused by the wide range of
formation histories and galaxy types responsible for the DLAs.  In
fact, Pettini et al. (1999) have also concluded that the known DLAs
are unlikely to trace the galaxy population responsible for the bulk
of the star formation.

Prochaska \& Wolfe (1997; 1998) have used high-resolution spectra of
the metal-line profiles associated with DLAs to argue that the
kinematic profiles of the DLA absorbing regions are more consistent
with models of rotating HI disks than any other single type of model.
The kinematics have also been shown to be consistent with gas in-fall
due to merging (Haehnelt, Steinmetz, \& Rauch 1998), randomly moving
clouds in a spherical halo (McDonald \& Miralda-Escud\'e 1999), and
multiple gas disks in a common halo (Maller et al.  2000). Therefore,
it seems most likely that, consistent with other current findings, the
kinematics of DLA absorbing regions arise from a mix of kinematic
structures, including some rotating gaseous disks.

There is now direct evidence from imaging studies that the
morphological types of DLA galaxies are indeed mixed.  See Rao \&
Turnshek (1998), Turnshek et al.  (2001), and Nestor et al. (2001,
these proceedings) for ground-based imaging results on some of our
low-redshift DLA fields.  HST images of other low-redshift DLA
absorbers have also revealed a mixed population (Le Brun et al. 1997;
Bouch\'e et al. 2001). Turnshek, Rao, \& Nestor (2001, these
proceedings) compare the properties of low-redshift DLA galaxies  to
the properties of gas-rich galaxies at $z=0$.

Thus, there has been a shift in our interpretation of the nature of
the DLA galaxy population, consistent with the idea that DLAs arise in
giant hydrogen clouds that could be associated with any type of galaxy
or protogalaxy (Khersonsky \& Turnshek 1996).  Here, we elaborate on
some of the  statistical results from our DLA survey.

\section{The Low-Redshift DLA Survey}

We recently completed an efficient non-traditional (but unbiased) HST
survey to discover DLAs at redshifts $z<1.65$ (RT2000).  Our survey
relied on observations of Ly$\alpha$ absorption in identified MgII
systems (which can be identified from the ground for $z>0.1$), where
the incidence of MgII is known as a function of the MgII$\lambda$2796
rest equivalent width, \w\ (e.g. Steidel \& Sargent 1992).  The
empirical fact that all DLA absorbers have MgII absorption (Turnshek
et al. 1989; Lu et al.  1993; Wolfe et al.  1993; Lu \& Wolfe 1994)
greatly improves the efficiency of searches for DLA at low redshift.
With our survey technique, we uncovered 12 DLA lines in 87 MgII
systems with \w\ $\ge0.3$ \AA\ (a success rate of $\approx$14\%).  Two
more DLAs were discovered serendipitously.  In total our survey
increased the number of known low-redshift DLA absorbers more than
two-fold (previously known low-redshift DLAs were mostly identified
via 21 cm absorption).

\subsection{The Statistical Properties of Low-Redshift DLAs}

Since we observed the fraction of MgII systems with DLA, we determined
the incidence of DLAs at low redshift by bootstrapping from the MgII
statistics.  By fitting Voigt damping profiles to the Ly$\alpha$ lines
in our UV spectra, we deduced $N_{HI}$ for each DLA system and
determined the low-redshift cosmological neutral gas mass density of
DLA absorbers.  Specifically, our survey results indicate:

(1) Approximately 50\% of the systems with \w\ $\ge$0.5 \AA\ {\it and}
FeII $W_0^{\lambda2600}$ $\ge$0.5 \AA\ have DLA absorption (Figure 1).
Thus, we uncovered a new selection criterion for the identification of
DLAs. Moreover, all the non-DLAs in this regime have HI column
densities $N_{HI}\ge 10^{19}$ atoms cm$^{-2}$.  We also found that
with the exception of the known $z=0.692$ 21 cm absorber towards 3C
286 which has unusually low metal-line equivalent widths (\w\ $=0.39$
\AA\ and $W_0^{\lambda2600}=0.22$ \AA, Cohen et al. 1994), all of the
DLAs lie in the upper-right region of the plot where \w\ $>0.5$ \AA\
and $W_0^{\lambda2600}>0.5$ \AA.

\begin{figure}
\plotfiddle{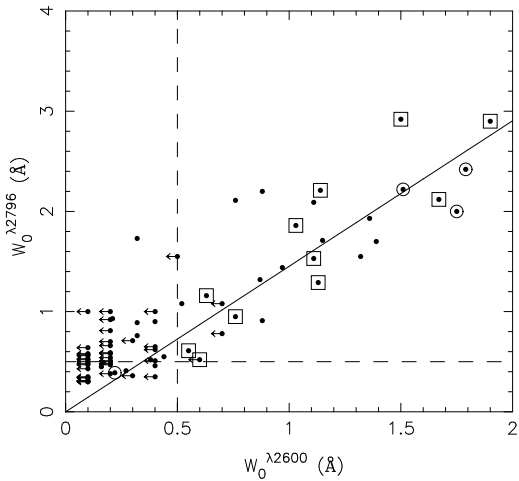}{6.0cm}{0}{50}{50}{-150}{-140}
\caption{\footnotesize A plot of MgII \w\ versus FeII
$W_0^{\lambda2600}$.  The DLAs from the RT2000 sample are marked with
open squares.  The open circles represent previously known 21 cm
absorbers that were excluded from our unbiased MgII sample.  Left
pointing arrows indicate upper limits to the measured value of
$W_0^{\lambda2600}$.  The horizontal and vertical dashed lines
identify the region for which \w$>0.5$ \AA\ and
$W_0^{\lambda2600}>0.5$ \AA; half of these are DLAs. }
\end{figure}

(2) The incidence of DLAs per unit redshift, $n_{DLA}$, is observed to
decrease with decreasing redshift (Figure 2).  The observed trend in
$n_{DLA}$ implies significant evolution only if the $z=0$ data point,
which is derived from 21 cm observations of gas-rich spirals (Rao \&
Briggs 1993), is included in the analysis. Two power laws of the form
$n_{DLA}(z)=n_0 (1+z)^\gamma$ are shown. For $\Lambda=0$ cosmologies,
$n_{DLA}$ is consistent with no intrinsic evolution if $\gamma=1.0$
for $q_0=0$ or if $\gamma=0.5$ for $q_0=0.5$. The solid line in Figure
2, which goes through the $z=0$ data point, has exponent $\gamma=2.5$
and indicates significant evolution.  The dashed line, which excludes
the $z=0$ data point, has $\gamma=1.5$ and is only marginally
consistent with evolution.  Extrapolation to $z=0$ (the open triangle)
results in a value that is $\sim3$ times larger than the observed
incidence at $z=0$.

\begin{figure}
\plotfiddle{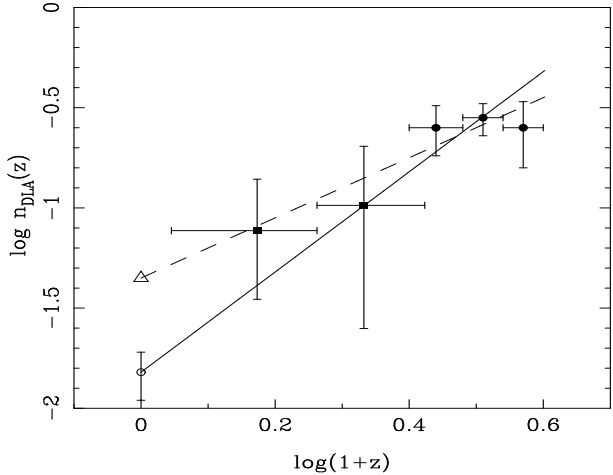}{6.0cm}{0}{60}{50}{-190}{-130}
\caption{\footnotesize The number density redshift distribution of
DLAs.  Our low-redshift results are shown as solid squares (RT2000).
A Malmquist bias correction factor has been applied to the
high-redshift data of Wolfe et al. (1995) and the results are shown as
solid circles.  See RT2000 for details. The open circle is derived
from the observed HI distribution in local spiral galaxies (Rao \&
Briggs 1993).  The solid line has $\gamma=2.5$ and is forced to pass
through the $z=0$ data point, the dashed line has $\gamma=1.5$ and
does not include the $z=0$ data point. Extrapolation of this power-law
to $z=0$ (the open triangle) results in a value that is $\sim3$
times larger than the observed incidence at $z=0$. }
\end{figure}

(3) The cosmological mass density of neutral gas in low-redshift DLA 
absorbers, $\Omega_{DLA}$, is found to be comparable to that observed
at high redshift (Figure 3).  The error bars are large because
$\Omega_{DLA}(z)$ is very sensitive to the small number of systems
which have the highest column densities. For $q_0=0.5$, $\Omega_{DLA}(z)$
is a factor of $\sim6.5$ times larger than the value for $\Omega_{gas}(z=0)$ 
inferred from local 21 cm observations. 

\begin{figure}
\plotfiddle{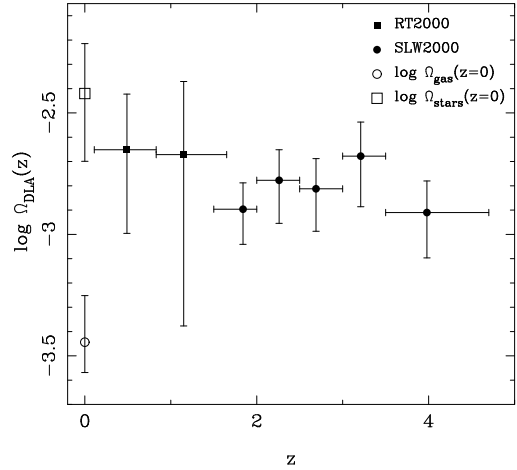}{6.0cm}{0}{50}{50}{-150}{-130}
\caption{\footnotesize The cosmological mass density of neutral gas as
a function of redshift for $H_0=65$ km s$^{-1}$ Mpc$^{-1}$, $q_0 = 0.
5$, and $\Lambda=0$.  The data points for $z>1.5$ are from
Storrie-Lombardi and Wolfe (2000).  The open circle at $z=0$ is the
local neutral gas mass density as measured by Rao \& Briggs (1993) and
the open square at $z=0$ is the local luminous mass density in stars
(Fukugita, Hogan, \& Peebles 1998).}
\end{figure}

(4) The HI column density distribution (CDD) of the low-redshift DLA
absorber population is very different in comparison to high-redshift
DLA absorbers, and in comparison to the column density distribution
inferred from local spirals (Figure 4). The low-redshift DLAs
exhibit a significantly larger fraction of very high column density
systems in comparison to determinations at both high redshift and
locally.  At no redshift does the CDD fall-off in proportion to
$\sim$$N_{HI}^{-3}$. An $\sim$$N_{HI}^{-3}$ fall-off is theoretically
predicted for disk-like systems (e.g.  Milgrom 1988) and this is, in
fact, observed locally in spiral samples (Rao \& Briggs 1993; Zwaan et
al. 1999).

\begin{figure}
\plotfiddle{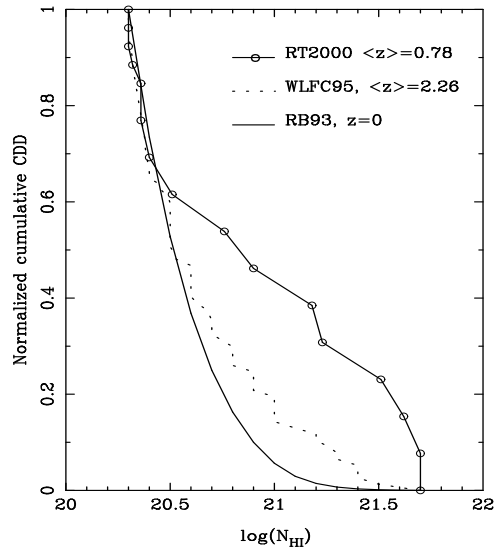}{6.0cm}{0}{70}{60}{-250}{-150}
\caption{\footnotesize The normalized cumulative column density
distributions for the low-redshift DLA sample (RT2000),
the high-redshift DLA sample (Wolfe et al. 1995), and local galaxies
(Rao \& Briggs 1993).}
\end{figure}

Despite the increase in the number of known low-redshift DLAs, the
statistics are still dominated by small numbers leading to large error
bars on $n_{DLA}$ and $\Omega_{DLA}$.  It could be argued that these
quantities are also consistent with a real decrease between
$z\approx4$ and $z\approx0.5$. However, the CDD at low redshift is
very significantly different from that at high redshift. There is only
a 2.8\% probability that the two samples are drawn from the same
parent population. Also, the low-redshift DLA sample is different from
the local spiral galaxy population at the 99.99\% confidence level.

\section{Discussion}

In summary, the observational results on DLAs over the past decade
have shown that their redshift number density, cosmological mass
density, and metallicities show little evolution from redshifts
$z\approx4$ to $z\approx0.5$. We have also seen that DLA galaxies
cover a wide range of gas-rich morphological types from predominantly
dwarf and low surface brightness galaxies to some spirals.

That they are a slowly evolving population implies that either: (1)
they do not trace the bulk of the galaxy population responsible for
star formation, i.e., they are a different population than the
galaxies that are used to determine the star formation history of the
Universe, or (2) that the galaxy population responsible for star
formation is a small subset of the DLA galaxy population. If the first
possibility holds, this is probably the result of a selection effect
in that optical surveys of QSOs preferentially probe lines of sight
that are relatively dust free. These sight lines must avoid star
forming regions that are enshrouded in metal-enriched dust (Pei, Fall,
\& Hauser 1999). Indeed, the models of Pei, Fall, \& Hauser (1999)
imply that as much as $\sim$70\% of the neutral gas mass is being
missed by DLA surveys. However, the recent study of DLAs detected in a
sample of  radio-selected quasar spectra by Ellison et al. (2001)
suggests that  the effect might be less important.

Our results on the incidence of DLAs show that $n_{DLA}$ is only
marginally consistent with evolution from $z\approx4$ to $z\approx
0.5$, and  that the extrapolated value of $n_{DLA}(z)$ at $z=0$  is
higher than that deduced from 21 cm emission data of nearby galaxies
by a factor of $\sim3$ (see Figure 2). We also found that
$\Omega_{DLA}$ does not evolve from $z\approx4$ to $z\approx0.5$ and
that the extrapolated value at $z=0$ is  a factor of $\sim$6.5
times higher than the neutral gas mass density deduced from gas-rich
galaxies at the present epoch (Figure 3). Thus, the QSO
absorption-line  results are highly inconsistent with the results from
21 cm observations  of local galaxies. Recently, Churchill (2001) used
HST archival spectra to determine the incidence of MgII systems at
$z\approx 0.05$ and, based on the RT2000 method of bootstrapping from
MgII statistics and the assumption  of no evolution in the MgII-to-DLA
statistics,  derived a value for $n_{DLA}(z\approx0.05)$ which is
essentially equivalent to the RT2000 value for
$n_{DLA}(z\approx0.5)$. He further noted that if the CDD of DLAs did
not evolve from $z\approx0.5$ to $z\approx0.05$, then
$\Omega_{DLA}(z\approx0.05)$ would  also be consistent with the RT2000
result for $\Omega_{DLA}(z\approx0.5)$. The implication, then, would
be  that $\Omega_{DLA}$ is constant from $z\approx 4$ to nearly the
current epoch, i.e., $z\approx0.05$. If the Churchill (2001) results
are confirmed  with follow-up observations of the corresponding
Ly$\alpha$ lines then the discrepancy between the QSO absorption line
results and 21 cm emission measurements of local galaxies would have
to be explained.

\subsection{At low redshift}

It is possible that we are so dominated by errors from the statistics
of small numbers, that our sample just happened to have a higher
fraction of the highest column density systems by sheer chance. As
mentioned above, the Churchill (2001) result on
$\Omega_{DLA}(z=0.05)$ was derived assuming that the CDD of DLAs did
not evolve from $z\approx0.5$ to $z=0.05$. So that result is also
subject to question in the same way.  The only way around this  is to
conduct larger surveys for DLAs at low redshift. In addition to
improving the statistical uncertainties, this would permit a better
determination of the shape of the HI CDD at large column densities,
effectively allowing a determination of the maximum HI column density.
Understanding this turn-down in column density is important for the
determination of $\Omega_{DLA}$.  It is also worth reiterating that
the redshift interval $0<z<1.65$ includes the most recent $\sim$70\%
of the age of the Universe. We certainly require more than a dozen or
so DLAs to do justice to this era.

It might be argued that our technique of selecting from a MgII sample
biases the DLA  sample towards higher column densities. However, we
have shown that our data are not consistent with this (see RT2000).
There is no trend between MgII \w\ and HI column density. While we
used a \w\ detection threshold of 0.3\AA, all but one of our DLAs
have \w\ $>0.6$\AA\ (the one system has \w\ $=0.52$\AA).  Thus, we
believe that our empirically-determined cut-off in \w\ is sound. It
was also suggested by Frank Briggs at this conference that a higher
\w\ cut-off should be used to search for DLAs since they are probably
more indicative of the presence of a DLA system. Moreover, the Steidel
\& Sargent (1992) MgII study showed that systems with \w\ $>1.0$ \AA\
evolved away faster than those with \w\ $>0.6$\AA. This might, he
argued, result in a smaller value for $\Omega_{DLA}(z\approx0.5)$. We
find that the results on $\Omega_{DLA}(z\approx0.5)$ from the two \w\
threshold samples in our data set are consistent with each other; of
course, the error bars for the  \w\ $>1.0$\AA\ sample are even larger.

It is also a concern that our QSO samples might be influenced by
gravitational lensing bias; DLA galaxies magnify QSOs,  preferentially
introducing them into optically selected samples.  Le Brun et
al. (2000) have shown this effect to be less than 0.3 magnitudes for
their sample of 7 DLA galaxies. They also point out that our QSO
sample is brighter than theirs and that magnification bias may,
therefore, be more important for our sample. The QSOs with  the
highest column density DLAs in our sample have not been imaged  with
HST, and so the possibility of lensing by the DLA galaxies has not
been studied with the greatest possible sensitivity. On the other
hand, our ground-based studies suggest that DLA galaxies are not
highly luminous, and therefore, massive galaxies.

In any case, it is clear that in addition to increasing DLA sample
sizes, biases due to MgII selection, dust obscuration, and the effect
of gravitational lensing could be assessed more carefully.

\subsection{At the present epoch}

We should also re-examine the $z=0$ statistics. In Rao \& Briggs
(1993) we used the best available optical luminosity  functions known
at that time in conjunction with empirical HI mass -- optical
luminosity relations of gas-rich galaxies to calculate the HI mass
density locally. This is the value plotted in Figure 3. The upper
error bar has been modified slightly as explained in Rao, Turnshek, \&
Briggs (1995). Fall \& Pei (1993) derived approximately the same value
using mean values of the local luminosity density and $M_{HI}/L_B$.
This exercise was repeated by Natarajan \& Pettini (1997) using  more
recent optical luminosity functions and they confirmed the Rao \&
Briggs (1993) result.  For comparison, the results on
$\Omega_{gas}(z=0)$ were $2.4\times10^{-4}h^{-1}$ (Rao \& Briggs
1993),  $2.6\times10^{-4}h^{-1}$ (Fall \& Pei 1993), and
$2.5\times10^{-4}h^{-1}$ (Natarajan \& Pettini 1997).

Direct measurements of the local  gas mass density using HI 21 cm
emission surveys have also been found to be consistent with the  above
results.  The Arecibo HI Strip Survey (AHISS, Zwaan et al. 1997)
performed the most sensitive search for local 21 cm emission to date,
being 5 times more sensitive than the Arecibo Dual Beam Survey (ADBS)
of Rosenberg \& Schneider (2000), although the latter  covered a
larger volume. The limiting column density of the Zwaan et al. survey
was $10^{18}$ cm$^{-2}$ at the $5\sigma$ level for  gas filling the
telescope beam, and they had the capability of  detecting HI masses of
$6\times10^5 h^{-2}$M$_{\odot}$ at 7$h^{-1}$ Mpc. For comparison,
HIPASS, which is a blind 21 cm survey of the  southern sky, reached a
limiting column density of $7\times 10^{17}$ cm$^{-2}$ at the
$3\sigma$ level and a mass limit of  $\sim7\times10^7h^{-2}$
M$_{\odot}$ at 7$h^{-1}$ Mpc (Kilborn, Webster, \& Staveley-Smith
1999; Kilborn 2001). All three 21 cm surveys have been used to
construct the local HI mass function. The faint end slope of the HI
mass function has been the focus of much debate since it bears
directly on the values of  $\Omega_{gas}(z=0)$ and  $n(z=0)$ (Figures
2 and 3). Moreover, a large faint end slope might imply the existence
of a new dwarf galaxy population that has not been identified
optically. For comparison,  Zwaan et al. (1997) derive a slope of
$\alpha=1.2$, Rosenberg \& Schneider (2001, these proceedings) 
get $\alpha=1.5$, while
Kilborn (2001) derives $\alpha=1.52$ (where $\alpha$ is the standard
Schechter luminosity function parameter).  The integral of the HI mass
function is an estimate of the total HI mass density at $z=0$, and its
ratio with the critical mass  density at the current epoch gives
$\Omega_{HI}(z=0)$. A correction for a neutral gas composition of 75\%
H and 25\% He by mass then gives $\Omega_{gas}(z=0)$.  The results on
$\Omega_{gas}(z=0)$ for the three surveys  are $2.7\times 10^{-4}
h^{-1}$ (Zwaan et al. 1997), $3.8 \times 10^{-4} h^{-1}$ (Kilborn
2001), and  $4.4 \times 10^{-4} h^{-1}$ (from data in Rosenberg \&
Schneider 2001, these proceedings).

Thus, the value of $\Omega_{gas}(z=0)$ derived from  optical
luminosity functions is consistent with the value derived  from HI
mass functions. As expected, the HI mass functions with  $\alpha\sim
1.5$ result in slightly larger values for  $\Omega_{gas}(z=0)$, but
only at the $\sim2\sigma$ level.  The similarity between the HI mass
function estimated from AHISS  with that derived from optical
luminosity functions (Briggs \& Rao 1993) led Zwaan et al. (1997) to
conclude that there is no large population of HI-rich galaxies
(e.g. low surface brightness or dwarf) that have been missed by
optical galaxy surveys. See also Zwaan, Briggs, \& Sprayberry (2001)
who show that the optical luminosity function of HI selected galaxies
is in agreement with the luminosity function of optically selected
late-type galaxies. However, based on the sensitivities of  the three
21 cm surveys discussed above, the possibility of there being objects
with masses lower than $\sim 10^7$ M$_{\odot}$ that contain DLA column
densities, but that have a mean column density less  than $\sim
10^{18}$ cm$^{-2}$ within the telescope beam cannot be ruled out.

Rosenberg \& Schneider (2001, these proceedings)  have also suggested
that the faint-end slope of the ADBS HI mass function, $\alpha=1.52$,
is large enough to make the contribution of low-HI-mass galaxies
important to the DLA cross-section at $z=0$. They derive a value for
$n(z=0)$ that is a factor of $\sim5.5$ higher than our estimate from a
complete sample of 27 gas-rich galaxies (see Rao, Turnshek, \& Briggs
1995 and Turnshek, Rao, \& Nestor 2001, these proceedings).  While the
ADBS sample covers a larger range in HI masses ($5\times10^7$M$_\odot$
to $2\times 10^{10}$M$_\odot$) compared to our sample
($5\times10^8$M$_\odot$ to $2\times10^{10}$M$_\odot$), the
cross-sectional areas of our galaxies at the $2\times10^{20}$ atoms
cm$^{-2}$ HI column density contour  are systematically lower by a
factor of $\sim 4$.  This might possibly account for the larger value
of $n(z=0)$ that they derive from their data set.

\section{Concluding Remarks}

Observations over the past several years have led us to a new paradigm
for DLAs, namely, that DLAs trace clouds  of neutral gas whose
properties have not changed significantly between $z\approx4$ and
$z\approx0.5$, and that these clouds  arise in a variety of gas-rich
galaxy types. There exists a  discrepancy of about a factor of 6.5
between the cosmological  neutral gas mass density at $z=0$ as
inferred from DLA statistics  and 21 cm emission measurements of local
galaxies. Whether this  discrepancy is due to small-number statistics
of DLAs at low redshift, a real selection  difference, or real
evolution can only be resolved with more and  better data at low
redshift and at $z=0$.

This research has been supported by STScI grant GO-06577.01-95A and
NASA LTSA grant NAG5-7930.

\end{document}